\documentclass[twocolumn,prl,showpacs,superscriptaddress,amssymb,amsmath,amsmath]{revtex4-1}
\usepackage{graphicx}
\usepackage{epstopdf}
\usepackage{bm}
\usepackage{hyperref}
\usepackage{comment}

\hypersetup{%
   pdfpagemode=None, 
   pdfstartpage=1,
   pdfmenubar=true,
   pdftoolbar=true,
   colorlinks = true,
   linkcolor=blue,
   citecolor=blue,
   urlcolor=blue,
   bookmarksopen=false
 }

\begin{document}
\title{Energetics and control of ultracold isotope-exchange reactions \\between heteronuclear dimers in external fields}

\author{Micha\l~Tomza}
\email{michal.tomza@chem.uw.edu.pl}
\affiliation{ICFO-Institut de Ci\'encies Fot\'oniques, Av.~Carl Friedrich Gauss, 3, 08860 Castelldefels, Spain}
\affiliation{Faculty of Chemistry, University of Warsaw, Pasteura 1, 02-093 Warsaw, Poland}

\date{\today}

\begin{abstract}

We show that isotope-exchange reactions between ground-state alkali-metal, alkaline-earth-metal, and lanthanide heteronuclear dimers consisting of two isotopes of the same atom are exothermic with an energy change in the range of 1-8000$\,$MHz thus resulting in cold or ultracold products. For these chemical reactions there are only one rovibrational and at most several hyperfine possible product states. The number and energetics of open and closed reactive channels can be controlled by the laser and magnetic fields. We suggest a laser-induced isotope- and state-selective Stark shift control to tune the exothermic isotope-exchange reactions to become endothermic thus providing the ground for testing models of the chemical reactivity. The present proposal opens the way for studying the state-to-state dynamics of ultracold chemical reactions beyond the universal limit with a meaningful control over the quantum states of both reactants and products.   
 
\end{abstract}


\maketitle

Ultracold molecules have recently become a new powerful tool for investigating chemistry at its fundamental quantum limit~\cite{ColdMolecules}. The production of an ultracold high phase-space-density gas of polar molecules in their absolute rovibrational ground state~\cite{NiScience08,TakekoshiPRL14,MolonyPRL14} allowed for groundbreaking experiments on controlled chemical reactions~\cite{OspelkausScience10,NiNature10,MirandaNatPhys11}. An unprecedented control over the quantum states of reactants both by selecting their internal states and by tuning dipolar collisions with an external electric field in a reduced dimensionality were achieved. The next expected milestone towards the state-to-state ultracold controlled chemistry will be the measurement and control of the product-state distributions~\cite{KremsPCCP08}.  

The quantum statistics~\cite{OspelkausScience10}, quantum tunneling and reflection from a centrifugal barrier~\cite{IdziaszekPRL10}, as well as the impact of external fields~\cite{MirandaNatPhys11} can modify the rates for ultracold chemical reactions by many orders of magnitude because at ultralow temperatures the reactants can be prepared in a single quantum state and even a weak perturbation can be comparable to or larger than the collision energy~\cite{KremsPCCP08}. At the same time, the energy released in a chemical reaction between $^1\Sigma^+$-state polar alkali-metal molecules in the ground rovibrational level, if energetically allowed, is at least 12$\,$K for KRb, or more for other molecules~\cite{ZuchowskiPRA10}. This means that untrappable products in many states with relatively large kinetic energy are expected and thus prospects for observing quantum effects and control in product distributions are limited~\cite{GonzalezPRA14}.

The rate constants for ultracold reactive collisions of the KRb molecules were shown to be universal, that is, independent of the short-range dynamics when all wave function amplitude transmitted to a short range is lost~\cite{IdziaszekPRL10}. To avoid reactive losses, which could obstruct the interesting applications in quantum simulations of many-body physics~\cite{BoNature13}, the creation of polar alkali-metal molecules in the rovibrational ground state stable against chemical reactions emerged as another important research goal~\cite{TakekoshiPRL14,MolonyPRL14}. 
However, recently it was pointed by Bohn and collaborators~\cite{MaylePRA12,MaylePRA13,CroftPRA14} that even chemically stable molecules can form long-lived collision complexes which can lead to the delayed three-body loss mechanism. This prediction based on the statistical approach together with other open questions on the dynamics of the reaction complex, non-universal reactivity, and potential control of state-to-state transitions
still require  experimental verification.
For the systematic investigation of ultracold chemical reactions, model systems with a small number of entrance and exit channels, and significant control over the quantum states of both reactants and products are of great importance.
 
Here we propose ultracold isotope-exchange reactions between heteronuclear dimers---a prototype of atom-exchange reactions---as a promising system for investigating models of the controlled chemical reactivity. We show that the isotope-exchange reactions between all heteronuclear alkali-metal, alkaline-earth-metal, and lanthanide dimers are exothermic with the reaction energy change as small as 1$\,$MHz thus resulting in only one rovibrational and at most several hyperfine possible product states with the kinetic energy in the cold or ultracold regime. We demonstrate that these exothermic chemical reactions can be tuned to become endothermic by employing light near-resonant with a molecular bound-to-bound transition in the heteronuclear reactant while off-resonant for homonuclear products, that is, isotope- and state-selective AC Stark shift control. Application of the magnetic field can bring an additional control over the number and energetics of open and closed reactive channels. The present proposal thus paves the way for studying ultracold chemical reactions beyond the universal limit with a meaningful control over the quantum states of both reactants and products.

The AC Stark shift control over the entrance channels and reaction complex was recently proposed for a bimolecular reaction~\cite{WangPRL14} and experimentally realized for unimolecular~\cite{SussmanScience06} and atomic~\cite{BauerNatPhys09} processes. 


\paragraph{Isotope-exchange reactions.---} We consider chemical reactions between the ground-state heteronuclear dimers consisting of two isotopes of the same atom
\begin{equation}\label{eq:reaction}
{}^nA{}^mA+{}^nA{}^mA\to{}^nA_2+{}^mA_2\,.
\end{equation}
They will lead to collisional losses only when energetically allowed. Within the harmonic oscillator approximation used to describe the rovibrational ground states, the energy change for the reaction~\eqref{eq:reaction} is given by: 
\begin{equation}\label{eq:dE}
\Delta E_\mathrm{ex}\approx-\frac{\omega_0(M_m-M_n)^2}{128\mu^2}\,,
\end{equation}
where $\omega_0$ is the harmonic frequency of the ground state, $\mu$ is the reduced mass of the ${}^mA{}^nA$ molecule, and $M_m-M_n$ is the mass difference of the isotopes ${}^mA$ and ${}^nA$.
It means that all isotope-exchange reactions of the type \eqref{eq:reaction} are exothermic and all heteronuclear dimers are chemically unstable at ultralow temperatures. The energy released in the process is much smaller than the vibrational spacing, $\Delta E_\mathrm{ex}\ll\omega_0$, therefore the products are possible only in their vibrational ground state. If the reaction change is smaller than the lowest  rotational excitation energy, $\Delta E_\mathrm{ex}<2B_0$ ($B_0$ is the rotational constant), then products are possible only in their rovibrational ground state. 
Equation \eqref{eq:reaction} predicts that smaller reaction energy is expected for heavier dimers with smaller harmonic frequency and smaller mass difference of involved isotopes.

To give an accurate prediction of the energy change for the isotope-exchange reactions we have performed full calculations for selected alkali-metal and alkali-earth-metal dimers by subtracting the rovibrational energies of the ground-state reactants from those of products. The rovibrational energy levels have been calculated from the available accurate experimental potential energy curves of the $X^1\Sigma^+$ electronic ground state of the Li$_2$~\cite{SemczukPRL14}, K$_2$~\cite{PashovEPJD08}, Rb$_2$~\cite{StraussPRA10}, Sr$_2$~\cite{SteinPRA08} molecules and of the $a^3\Sigma^+$ state of the Li$_2$~\cite{SemczukPRL14}, K$_2$~\cite{PashovEPJD08}, Rb$_2$~\cite{StraussPRA10} molecules. 
The hyperfine structure and mass-dependent adiabatic and QED corrections to the Born-Oppenheimer interaction potentials are negligible~\cite{AldegundePRA08}.
The expression~\eqref{eq:dE} reproduces the exact values within only a few percent error, therefore the reaction energy changes for the Yb$_2$, Dy$_2$, Eu$_2$ dimers, for which the experimental data are not available, were obtained with this equation and theoretically predicted harmonic frequencies~\cite{BuchachenkoEPJD07,BuchachenkoJCP09,KotochigovaPRL12}.

The energies released in the isotope-exchange reaction between dimers in the rovibrational ground level of the $^1\Sigma^+$ electronic state are presented in Table~\ref{tab:dE} and take values between 0.87~MHz for $^{173}$Yb$^{174}$Yb up to 8104~MHz for $^6$Li$^7$Li. For $^{85}$Rb$^{87}$Rb and $^{87}$Sr$^{88}$Sr they are 29.1~MHz and 5.0~MHz, respectively. 
The formation of products is energetically allowed only in the rovibrational ground state.
The energy of 20~MHz corresponds to 1~mK, therefore the products of all considered reactions have
kinetic energy in the cold or ultracold regime.
If the external magnetic field can be used to restrict the total electron spin of the collision complex to its maximum value~\cite{TomzaPRA13b}, then the low-energy isotope-exchange reactions can be investigated between alkali-metal dimers in the $^3\Sigma^+$ state, for which the energy change is smaller by a factor of 4-5 as compared to the values for the $^1\Sigma^+$ state [cf.~Table~\ref{tab:dE}].

\begin{table}[t!]
\caption{The energy change $\Delta E_\mathrm{ex}$ for the reactions $2{}^mA{}^nA\to {}^mA_2 + {}^nA_2$ of heteronucelear dimers in the rovibrational ground state of the $X^1\Sigma^+$ or $a^3\Sigma^+$ electronic state and the estimated Rabi frequency $\Omega$ of the laser field needed to suppress reactivity ($\Delta E_\mathrm{AC}=\frac{1}{2}\Delta E_\mathrm{ex}$) assuming the detuning $\Delta^{nm}=10\Gamma_e$. $B_0$ and $\omega_0$ are the rotational constants and harmonic frequencies, respectively, for the rovibrational ground state of the ${}^mA{}^nA$ dimer.}\label{tab:dE}
\begin{ruledtabular}
\begin{tabular}{lrrrr}
 dimer & $\Delta E_\mathrm{ex}$ (MHz) & $\Omega$ (MHz) & $B_0$ (MHz) & $\omega_0$ (GHz) \\
\hline
\multicolumn{5}{c}{$X^1\Sigma^+$} \\
\hline
${}^6$Li${}^7$Li         & -8104 & 8466 & 21700 & 10920 \\
${}^{39}$K${}^{40}$K     & -55.3 &  207 &  1660 &  2746 \\ 
${}^{39}$K${}^{41}$K     & -214  &  447 &  1640 &  2729 \\ 
${}^{85}$Rb${}^{87}$Rb   & -29.1 &  148 &   663 &  1720 \\
$^{87}$Sr$^{88}$Sr       & -4.98 &   55 &   527 &  1216 \\ 
$^{86}$Sr$^{88}$Sr       & -20.2 &  112 &   530 &  1219 \\
$^{84}$Sr$^{88}$Sr       & -83.2 &  240 &   536 &  1226 \\
$^{173}$Yb$^{174}$Yb     & -0.87 &   22 &   282 &   746 \\
$^{168}$Yb$^{176}$Yb     & -56.3 &  189 &   284 &   750 \\
${}^{163}$Dy${}^{164}$Dy & -0.90 &   24 &   287 &   769 \\
${}^{160}$Dy${}^{164}$Dy & -14.5 &   97 &   290 &   772 \\
${}^{151}$Eu${}^{153}$Eu &  -3.7 &   47 &   276 &   680 \\
\hline
\multicolumn{5}{c}{$a^3\Sigma^+$} \\
\hline
${}^6$Li${}^7$Li         & -1510 & 1843 &  8650 &  1992 \\
${}^{39}$K${}^{40}$K     & -12.7 &   96 &   770 &   625 \\ 
${}^{39}$K${}^{41}$K     & -49.1 &   194 &   761 &   622 \\ 
${}^{85}$Rb${}^{87}$Rb   & -6.87 &   72 &   315 &   401 \\
\end{tabular}
\end{ruledtabular}
\end{table}

\begin{figure}[t!]
\begin{center}
\includegraphics[width=0.95\columnwidth]{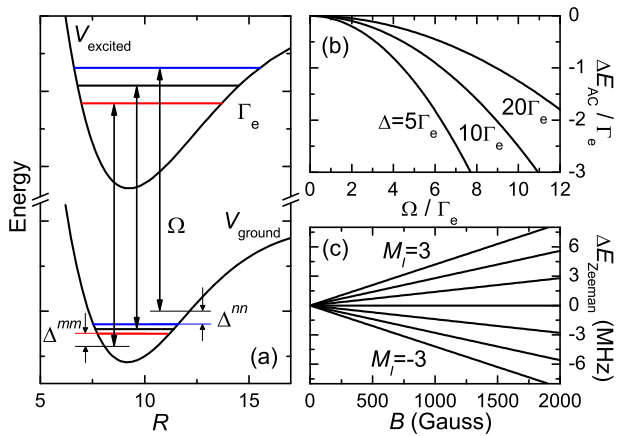}
\end{center}
\caption{(Color online) (a) Scheme of the state-selective AC Stark shift control. Different levels corresponds to different isotope mixtures (energies are not in scale). (b) The AC Stark shift for different laser detunings vs.~the laser field Rabi frequency $\Omega$. (c)~Zeeman levels of $^{87}$Rb$_2$ vs.~the magnetic field.}
\label{fig:Scheme}
\end{figure}


\paragraph{Stark shift control with laser field.---} 
Even small external perturbation can drastically modify the dynamics of the reaction~\eqref{eq:reaction} with such a small energy change. Unfortunately, the DC electric field control is not feasible for the ground-state homoatomic molecules. Here, we propose to use the isotope- and state-selective Stark shift induced by the cw laser field~\cite{AutlerPR55} to control the energetics of the reaction~\eqref{eq:reaction}.

The AC Stark shift for an effective two-level system in the weak-field regime ($\Omega\ll\Delta$) is given by:
\begin{equation}
\Delta E_\mathrm{AC}\approx\frac{\hbar\Omega^2\Delta}{4\Delta^2+\Gamma_e^2}\,,
\end{equation}
where $\Omega$ is the Rabi frequency of the laser, $\Delta$ is the detuning, and $\Gamma_e$ is the natural line width of the transition.
The proposed control scheme is based on finding the laser wavelength which is near-resonant with some bound-to-bound transition~\cite{BauerNatPhys09} in the heteronuclear reactant while off-resonant in the homonuclear products [cf.~Fig.~\ref{fig:Scheme}(a)].  
If $\Delta^{nm}\ll\Delta^{mm},\Delta^{nn}$, where $\Delta^{mn}$ is the detuning in the ${}^nA{}^mA$ molecule, then the significant AC Stark shift is experienced only by the heteronuclear dimer and modifies the energy change for the isotope-exchange reaction~\eqref{eq:reaction}. 
Potential heating and losses can be avoided by using the detuning much larger than the transition line width, $\Delta^{nm}\gg\Gamma_e$.
To employ the above control scheme in a realistic molecular system  the vibrational spacing for the ground and excited states must be larger than the detuning, $E_{v+1}^{nm}-E_{v}^{nm}\gg\Delta^{nm}$, and the excitation energy for different isotopomers must differ by more than the detuning, $|E_{v',v''}^{nn/mm}-E_{v',v''}^{nm}|\gg\Delta^{nm}$. 
The latter is guaranteed when the ground level is the vibrational ground state, whereas the excited level is the vibrational state from the middle of the spectrum, $v'\gg v''=0$.  
All these conditions can be met for the investigated dimers and only an accidental coincidence can result in the same Stark shifts for the hetero- and homonuclear dimers.

To ensure sufficiently large Rabi frequencies, we choose a transition to an excited state rovibrational level for which the atomic transition at the dissociation threshold is dipole allowed and strong~\cite{nist}. 
For alkali-metal atoms this is the $^2S_{1/2}-^2P_{3/2}$ transition at wavelengths from 670$\,$nm for Li to 780$\,$nm for Rb with the line width of 36-38$\,$MHz. For Sr and Yb, the $^1S_0-^1P_1$ transition at wavelengths 461$\,$nm and 399$\,$nm with the line widths of 30$\,$MHz and 29$\,$MHz, respectively.
For Dy and Eu, the $^5I_8-^5(8,1)_9$ and $^8S_{7/2}-^8P_{9/2}$ transitions at wavelengths 412$\,$nm and 466$\,$nm with the line widths of 32$\,$MHz and 30$\,$MHz, respectively. 

The molecular excitation energies and line widths will be different but of the same order of magnitude as the above listed atomic ones. Therefore, to estimate the Rabi frequency $\Omega$ of the laser needed to turn the exothermic isotope-exchange reactions into the endothermic ones,
we compare the AC Stark shift for the heteronuclear dimers with the reaction energy change,
$2\Delta E_\mathrm{AC}^{nm}(\Omega)=\Delta E_\mathrm{ex}$, assuming the detuning $\Delta^{nm}=10\Gamma_\mathrm{e}$. The obtained Rabi frequencies are collected in Table~\ref{tab:dE} and are between 22$\,$MHz for $^{173}$Yb$^{174}$Yb up to 447$\,$MHz for ${}^{39}$K${}^{41}$K and 8466$\,$MHz for ${}^6$Li${}^7$Li. The last one is quite large but for all other dimers the proposed AC Stark shift control should be experimentally feasible. 
 
To confirm the above predictions based on the simplified two-level calculations, we performed the full calculations of the AC Stark shift control by diagonalizing the molecular Hamiltonians for the ${}^{85}$Rb${}^{87}$Rb and $^{87}$Sr$^{88}$Sr dimers for which accurate electronic interaction potentials and transition dipole moments are known~\cite{StraussPRA10,SteinPRA08,TomzaMP13,SkomorowskiJCP12}.
For ${}^{85}$Rb${}^{87}$Rb (${}^{87}$Sr${}^{88}$Sr) we chose the laser wavelength $\lambda_L=619\,$nm (441$\,$nm) and the detuning $\Delta^{nm}=370\,$MHz (305$\,$MHz) from the level with the line width $\Gamma_e=37\,$MHz (30.5$\,$MHz). The Rabi frequency $\Omega=160\,$MHz (62$\,$MHz) is needed to suppress the reactivity. 
The calculations with the full Hamiltonians confirm the simplified predictions.


\paragraph{Control with magnetic field.---}
The $^1\Sigma^+$-state molecules have a zero electronic spin therefore the hyperfine splitting of the order of 1-100~kHz 
due to the nuclear spin-spin coupling occurs only in the case of two nuclei with a non-zero spin~\cite{AldegundePRA08}. If at least one nucleus in the heteronuclear reactant has a non-zero spin then products in a few hyperfine states are possible. Application of the magnetic field lifts the degeneracy of the hyperfine states and can be used to control energies of exit channels associated with different Zeeman levels. For example, the Zeeman shift for $^{87}$Rb$_2$ in the magnetic field of 1000$\,$Gauss is up to 4.2$\,$MHz [cf.~Fig.~\ref{fig:Scheme}(c)]. 


\paragraph{Applications.---} 
The proposed cold isotope-exchange reactions together with the control schemes provide a platform for testing models of the chemical reactivity.

The diversity of stable isotopes for Sr, Yb, or Dy can be used to tune the energy change, Eq.~\eqref{eq:dE}.
The atoms with several bosonic and fermionic isotopes give the unique opportunity for realizing isotope-exchange reaction between both fermionic and bosonic dimers and
 observing the effect of the quantum statistics.

If the reaction proceeds via the formation of an intermediate complex with a sufficiently long lifetime, whose dynamics renders the reactant and product channels statistically independent,
then the probability for the state-to-state transition $P_{\alpha\to\beta}$ (related to reaction rate) can be described within the statistical quantum model~\cite{MillerJCP70} by:
\begin{equation}
P_{\alpha\to\beta}\approx {p_\alpha p_\beta}/{\sum_\gamma p_\gamma}\,,
\end{equation}
where $p_i$ is the capture probability of the complex formation for the collision in the channel $i$.
In the limit of many more exit than entrance channels, the total rates for chemical reactions are predicted to depend solely on the capture dynamics in the reactant channels~\cite{GonzalezPRA14} and the universal limit of reactivity 
can be expected~\cite{IdziaszekPRL10}.

In the investigated systems, the rotational excitation can be used to control the number of exit rovibrational channels in the range from one to several by Raman excitation~\cite{DeissPRL14} or to several tens if excited by using an optical centrifuge~\cite{KorobenkoPRL14}.
This can allow to investigate the transition from the non-universal limit of the chemical reactivity to the universal one~\cite{JachymskiPRL13} in a single  experimental setup.  
The vibrational excitation can bring the system directly into the regime of the large number of exit channels and thus 
the universal limit of chemical reactivity~\cite{IdziaszekPRL10}. 

The reaction energy change can continuously be tuned from tens of MHz to zero with the proposed AC Stark shift control scheme.
This should allow to observe the impact of the quantum statistics, Wigner threshold laws, and quantum tunneling and reflection from the centrifugal barrier on the capture probabilities in the product channels and thus the quantum control of reaction rates by controlling the product states. 
The selective activation/deactivation of higher partial waves and tuning the position of shape resonances in the exit channels across the entrance threshold should also have a dramatic effect on the reaction rates and product distributions.
The estimated heights of the centrifugal barrier for the $p$-wave, $d$-wave, and $f$-wave collisions between  Sr$_2$ or Rb$_2$ dimers are 0.35$\,$MHz, 1.8$\,$MHz, and 5.2$\,$MHz, respecitively.

The comparison of the collisional losses when the chemical reactions are allowed with those when they are suppressed by the proposed control scheme should allow to verify the predictions based on the statistical models about the long-lived collision complexes formation and subsequent losses~\cite{MaylePRA12,MaylePRA13,CroftPRA14} in a single  experimental setup.
More specifically, it was predicted that the lifetime of the collision complex $\tau$ is proportional to the density of the available rovibrational states of the complex  $\rho$ and inversely proportional to the number of open channels $N_o$, $\tau\sim\rho /N_o$~\cite{MaylePRA13}. In the proposed experiments, the number of open channels can be controlled by rotational excitation or by using the combined Stark shift and magnetic field control with the number of open Zeeman levels from one to several.

The investigated reactions proceed barrierlessly via the four-body complex stabilized by non-additive forces~\cite{ZuchowskiPRA10,TomzaPRA13b}. The complex dynamics is still an open question and proposed experiments should also verify in a controlled way the statistical independence of the reactant and product channels and the existence of spectator degrees of freedom not involved during the reaction, that is, the conservation laws of different angular momenta present in reacting dimers.  
The control of the isotope-exchange reactions between the spin-polarized triplet-state dimers in a magnetic field can potentially give some insight into the dynamics of the transitions to lower spin states mediated by nonadiabatic spin-dependent couplings~\cite{JanssenPRL13}.


\paragraph{Experimental accesibility.---}
The preparation of ultracold dense gases of heteronuclear alkali-metal dimers in the temperature range of 1-100$\,\mu$K should not be more challenging than the formation of heteronuclear ${}^{40}$K$^{87}$Rb~\cite{NiScience08} and ${}^{87}$Rb$^{133}$Cs~\cite{TakekoshiPRL14,MolonyPRL14} molecules which were produced in their rovibration and hyperfine ground state by using the magnetoassociation followed by the efficient stimulated Raman adiabatic passage (STIRAP). 
Many other ultracold heteronuclear Bose-Bose, Fermi-Fermi, and Bose-Fermi mixtures of alkali-metal atoms were realized and converted into Fesbhach molecules~\cite{JulienneRMP10}, 
including ${}^{85}$Rb${}^{87}$Rb~\cite{PappPRL06}.
Actually, the formation of heteronuclear alkali-metal dimers can potentially be facilitated by similar trapping conditions and laser wavelengths needed for different isotopes as compared to different atoms.
The magnetoassociation of heteronuclear dimers from the highly magnetic lanthanide atoms has to face the problem of the high density of the interaction-anisotropy-induced Feshbach resonances~\cite{KotochigovaPRL12,FrischNature14} but then the STIRAP stabilization  should be feasible. 
The formation of heteronuclear molecules out of closed-shell atoms is a more challenging problem~\cite{TomzaPCCP11}. Nevertheless, the photoassocitive formation~\cite{JulienneRMP06a} of the homonuclear strontium~\cite{ReinaudiPRL12} and ytterbium~\cite{TojoPRL06}, and heteronuclear ytterbium~\cite{BorkowskiPRA11} dimers into weakly bound states were already demonstrated and a stabilization to the ground state can be expected.

The total reaction rates can be measured by monitoring the time evolution of the average number density of trapped heteronuclear reactants~\cite{OspelkausScience10,NiNature10,MirandaNatPhys11}. The state-selected product distribution measurement should also be possible for trappable ultracold homonuclear products~\cite{HarteNatPhys13}.
The estimated heating rate due to the photon scattering at a near-resonant transition used in the AC Stark shift control 
of the investigated Rb$_2$ and Sr$_2$ dimers is of the order of 1$\,$mK/s for $\Delta^{nm}=10\Gamma_e$ and can further be reduced by one to two orders of magnitude by increasing the used detuning $\Delta^{nm}$~\cite{GrimmAAMOP00}. This should be sufficiently low to allow for milliseconds measuraments. 

\paragraph{Summary.---}
We have shown that the ultracold isotope-exchange reactions between heteronuclear dimers in external fields provide the ground for testing models of the chemical reactivity where for the first time a meaningful control over the quantum states of not only reactants but also products can be realized.
These exothermic reactions can be tuned to become endothermic by employing the laser-induced isotope- and state-selective Stark shift control and the magnetic field brings an additional control over the number and energetics of open and closed reactive channels.
The demonstrated control schemes are also applicable to other isotope-exchange reactions e.g.~in ultracold mixtures of molecules and atoms. The proposed systems and required field strengths are all within the current experimental capabilities.

\begin{acknowledgments}
We would like to thank John Bohn, Christiane Koch, Robert Moszynski, and Jun Ye for useful discussions.
The author enjoyed the hospitality of the JILA at the University of Colorado at Boulder where the work
reported here was initiated. Financial support from the Foundation for Polish Science within the START program, the Marie Curie COFUND action through the ICFOnest program, the 
EU Grants OSYRIS, SIQS, QUIC, EQuaM, and the Spanish Ministry grant FOQUS   is gratefully acknowledged.

\end{acknowledgments}

\bibliography{../0Bib/MT}

\end{document}